# Macroeconomic factors and Stock exchange return: A Statistical Analysis

Md. Fazlul Huq Khan and Md. Masum Billah

## Abstract

The purpose of this research is to examine the relationship between the Dhaka Stock exchange index return and macroeconomic variables such as exchange rate, inflation, money supply etc. The long-term relationship between macroeconomic variables and stock market returns has been analyzed by using the Johnson Cointegration test, Augmented Dicky Fuller (ADF) and Phillip Perron (PP) tests. The results revealed the existence of cointegrating relationship between stock prices and the macroeconomic variables in the Dhaka stock exchange. The consumer price index, money supply, and exchange rates proved to be strongly associated with stock returns, while market capitalization was found to be negatively associated with stock returns. The findings suggest that in the long run, the Dhaka stock exchange is reactive to macroeconomic indicators.

## Introduction

The financial sector largely contributes to the economic growth and development of a country. The stock market is an integral part of the financial sector that provides necessary financial resources and support to investors. A well-functioning and efficient stock market provides the opportunity for companies to trade securities and provides valuable options for investments. Investors generate return from the dividends and capital gain by purchasing stocks of financially sound companies. The dividend yield and capital gain largely depend on the positive movement of the stock market. Due to macroeconomic factor changes, the stock market may depict volatile behaviour. Excessive volatility can increase the riskiness of the market and hinder the smooth functioning of the financial market. Investors may shift their investment to Govt. Securities and other risk-free assets. Identifying the dynamic and volatile behaviour of the stock market can reduce the risk to the investors and positively impact the financial sector.

Dhaka Stock Exchange (DSE) is the largest stock exchange in Bangladesh with over 350 companies and numerous treasury bills listed. DSE has experienced a couple of big shocks in its history and both times general investors had to suffer heavy losses. It took over a decade for the investors to recover from the financial suffering. Economic turmoil at national level can create strong volatility in the stock market. The economic factors that can affect the volatility of the stock market are Gross Domestic Product, inflation, money supply, interest rates, exchange rates, foreign direct investment, export growth etc. The relationship between the stock market and these economic factors can impact the share price.

## Literature Review

Zaheer & Rashid, (2014) investigated the relationship between Karachi stock market 100 index and macroeconomic variables, i.e., inflation, industrial production, money supply, exchange rate and interest rate. The authors identified the long-term relationship between macroeconomic variables and stock market returns by using Johnson Cointegration test, Augmented Dicky Fuller (ADF) and Phillip Perron (PP) tests. The Generalized Autoregressive Conditional heteroskedasticity (GARCH) model was used to find out the relationship between stock returns and the variance of the squared error terms as there was heteroskedastic trend in the data. The consumer price index (CPI), money supply (MS), exchange rates (ER) and interest rates (IR) proved to be negatively associated with the index returns whereas the industrial production index (IPI) was found to be positively associated with the stock market returns.

Acikalin et al. (2008) explored the relationship between macroeconomic variables and stock returns in Istanbul Stock Exchange (ISE) using quarterly time series data. The exogenous variables include Gross Domestic Product, interest rate, exchange rate and current account balance whereas Istanbul stock index was used as endogenous variable. Using Johnson Cointegration and Vector Error Correction model the authors found a long term and stable relationship between ISE index and macroeconomic variables. Gunsel & Cukur (2007) try to find out the impact of macroeconomic variables on the stocks returns of London stock exchange from 1980 to 1993. Interest rates, exchange rate, money supply, risk period, dividend yield etc variables are used against the stock returns. The investigation reveals that macroeconomic variables have positive and significant impact on the growth of stock index. Khan (2021) and Salma, (2021) reveals that exchange rate fluctuations and inflation can significantly affect the financial markets and trigger a downtrend in the economy.

Mukherjee and Naka (1995) analyzed the correlation between exchange rate, inflation, money supply, industrial production index, bond rate, call money rate and stock market index returns. Using vector error correction model the researchers there exists a long-term equilibrium relationship between the stock market returns and chosen macroeconomic variables. Ahmed (2021) tries to find the impact of unexpected events on stock market return and figured out that Covid-19 has severely impacted the stock market efficiency and stock returns. In another paper, Ahmed (2021) found that relationship between proper and transparent corporate governance and required disclosure to reduce agency conflicts and improve the financial performance of the companies that impacts the overall stock return of the market. Bhuiyan et al. (2020) also conducted the similar research and had the equivalent outcome.

Pal and Mittal (2011) investigated the association between Indian capital market returns and macroeconomic variables. Macroeconomic variables such as interest rates, inflation rate, exchange rates and gross domestic savings (GDS) of Indian economy were used as independent variables and two popular stock indices of India. Ahmed and Khan (2022) examined the impact of informational asymmetry on the return of the local loan markets. The data has been collected from 35 banks in 16 developing countries for 8 years period. The authors found that informational asymmetry can create a strong inefficiency in the investment market.

## Data and Methodology

Dhaka stock exchange index is the primary source of the data or the stock returns over the period. Macroeconomic variables information has been collected from the central bank publications and world bank database. All the monthly data for the year 2012-2022 with a total of 144 observations for each variable. Inflation measured as consumer price index (CPI), growth rate of money supply, exchange rate against U.S dollars and growth in market capitalization has been used as proxy for individual stock performance. The cointegration test suggested by Johnson (1990) has been used to overcome the problem of non-stationarity data sets. The Augmented Dicky Fuller (ADF) and Philip Perron (PP) unit root tests have also been applied to check the stationarity of the data. The regression model for this study is presented as follows:

$$DSE_{ij} = \alpha + \beta_1 EXH_{ij} + \beta_2 M2_{ij} + \beta_3 INF_{ij} + \beta_4 MCAP_{ij} + \varepsilon$$

The cointegration test involves the following equation

$$Y_j = \alpha_0 + \alpha_1 x_{1j} + \alpha_2 x_{2j} + \ldots\ldots\ldots + \alpha_k x_{kj} + \varepsilon$$

The cointegration equation shows the cointegrating relationship of all the independent variables with the dependent variable. In the cointegration equation, it is assumed that the each (k+1) series is stationary and homoscedastic.

**Table 1: Data Summary**

| Description | DSEX | EXH | M2 | INF | MCAP |
|---|---|---|---|---|---|
| Mean | -0.01208 | 4.41 | -0.00887 | 0.056698 | -0.00548 |
| Median | 0.003639 | 4.43 | -0.00728 | 0.05665 | -0.00482 |
| Minimum | -1 | 4.35 | -0.04396 | 0.035 | -0.1537 |
| Maximum | 0.086128 | 4.63 | 0.02 | 0.076 | 0.119311 |
| Std. Dev | 0.104505 | 1.50 | 0.009814 | 0.010485 | 0.04555 |
| Skewness | -8.19611 | 0.38 | -0.56525 | -0.35306 | -0.20502 |

**Estimation results**

Both the Augmented Dicky Fuller (ADF) and Philip Perron (PP) tests have been applied. The series are found to be difference stationary at level. However, the series are stationary at their first differences. Table 2 shows the results of both the Augmented Dicky Fuller and the Philip Perron unit root test at the level of the series.

**Table 2: Unit Root Test at First Differences**

| Variable | ADF t-statistics | PP t-statistics | Critical Value | | |
|---|---|---|---|---|---|
| | | | 1% | 5% | 10% |
| DSEX | -9.7423 | -9.7369 | -3.48 | -2.88 | -2.57 |
| EXH | -5.5411 | -5.6565 | -3.48 | -2.88 | -2.57 |
| M2 | -11.2347 | -11.2724 | -3.48 | -2.88 | -2.57 |
| INF | -8.8825 | -8.7831 | -3.48 | -2.88 | -2.57 |
| MCAP | -4.6616 | -5.2458 | -3.48 | -2.88 | -2.57 |

The Dhaka stock exchange index, exchange rate, money supply, consumer price index and market capitalization contain the T-statistics of -9.74, -5.54, -11.23, -8.88 and -4.66 respectively. The T-statistics for both the ADF and PP tests are less than the critical values.

**Table 3: Cointegration Test of Macroeconomic variables with DSEX**

| | No. of CE | Eigen Value | Trace Statistic | Prob |
|---|---|---|---|---|
| EXH | 1 | 0.1163 | 17.18 | 0.004 |
| M2 | 1 | 0.0977 | 19.21 | 0.000 |
| INF | 1 | 0.1021 | 28.21 | 0.011 |
| MCAP | 1 | 0.1578 | 15.14 | 0.068 |

Granger causality is a way to investigate causality between two variables in a time series. The method is a probabilistic account of causality; it uses empirical data sets to find patterns of correlation. The Table 4 shows the results of Granger causality test with the acceptance or rejection of the hypothesis of Granger causality.

**Table 4: Granger Causality Test**

| Hypothesis | Prob. | Decision |
|---|---|---|
| DSEX does not Granger Cause EXH | 0.0021 | Reject |
| DSEX does not Granger Cause M2 | 0.1610 | Accept |
| DSEX does not Granger Cause INF | 0.0030 | Reject |
| DSEX does not Granger Cause MCAP | 0.0000 | Reject |
| EXH does not Granger Cause DSEX | 0.0011 | Reject |
| EXH does not Granger Cause M2 | 0.0101 | Reject |
| EXH does not Granger Cause INF | 0.2354 | Accept |
| EXH does not Granger Cause MCAP | 0.0221 | Reject |
| M2 does not Granger Cause DSEX | 0.4710 | Accept |
| M2 does not Granger Cause EXH | 0.0364 | Reject |
| M2 does not Granger Cause INF | 0.0250 | Reject |
| M2 does not Granger Cause MCAP | 0.0361 | Reject |
| INF does not Granger Cause DSEX | 0.0261 | Reject |
| INF does not Granger Cause EXH | 0.3542 | Accept |
| INF does not Granger Cause M2 | 0.0142 | Reject |
| INF does not Granger Cause MCAP | 0.0607 | Reject |
| MCAP does not Granger Cause DSEX | 0.0587 | Reject |
| MCAP does not Granger Cause M2 | 0.0004 | Reject |
| MCAP does not Granger Cause EXH | 0.2550 | Accept |
| MCAP does not Granger Cause INF | 0.000 | Reject |

It has been observed that most of the macroeconomic variables Granger cause each other. The null hypothesis of no Granger causality has been rejected in case of most of the macroeconomic variables.

## Advanced Analytical Methods

Time Series reflects the sequence of data points in constant time intervals: daily, monthly, quarterly, yearly, and helps to track the changes over the predefined times. Time series analysis accounts for describing the internal structure of the sequential data and developing models which predict the future behavior of the dataset. After complex processing of current and historical data, time series forecasting builds the best-fit model to predict future trends most efficiently. Time series forecasting is a vital component of machine learning as it needs to consider lots of temporal components while processing. Roy et al. (2022) developed a machine learning time series methodology that can be used to predict the outcome of the exogenous variable based on the trends of endogenous variables. Pramanik, & Polansky, (2020) worked on a path integral model that can also work in time series structed and unstructured data. Moreover, nowadays machine learning algorithms have become more complex but efficient in predicting most challenging structured and unstructured time series issues.

## Conclusion

This research tries to identify the association between the macroeconomic variables with Dhaka stock exchange return. These macroeconomic variables are responsible for the variability in the financial market return and understanding the concepts are helpful to secure a steady return over time. The study finds that exchange rate fluctuations can create large volatility in the stock market. This variable is found to be significant and positively associated with stock returns. Money supply increases the demand for the stock and has a strong positive impact on the market return. Inflation reduces the purchasing power and therefore reduces the return. Inflation is found to have a negative but significant association with the stock exchange return. The market capitalization of the individual stocks has a minor but positive impact on the market returns. The financial market is crucial for the economy and any shock in the market can spill over the whole economy. The study can be helpful for the regulators, researchers and policy makers to identify and manage the volatility to generate a steady return for the investors.